\documentclass[twocolumn,showpacs,preprintnumbers,amsmath,amssymb]{revtex4}
\usepackage[english]{babel}
\usepackage[T1]{fontenc}
\usepackage[cp1250]{inputenc}
\usepackage{graphicx}
\usepackage{dcolumn}
\usepackage{bm}
\usepackage{ulem}
\newcommand{\sgm}{$\sigma^-$}
\newcommand{\sgp}{$\sigma^+$}
\begin{document}

\title{Probe spectroscopy in an operating magneto-optical trap:
\\the role of Raman transitions between discrete and continuum atomic states}

\author{Tomasz M. Brzozowski}
\email{tmb@ceti.pl}
\author{Maria Brzozowska}
\author{Jerzy Zachorowski}
\author{Micha{\l} Zawada}
\author{Wojciech Gawlik}
\affiliation{Marian Smoluchowski Institute of Physics,
Jagiellonian University, Reymonta 4, PL 30-059 Cracow}
\homepage{http://www.if.uj.edu.pl/ZF/qnog/}
\date{\today}

\begin{abstract}
We report on cw measurements of probe beam absorption and
four-wave-mixing spectra in a~$^{85}$Rb magneto-optical trap taken
while the trap is in operation. The trapping beams are used as pump
light. We concentrate on the central feature of the spectra at small
pump-probe detuning and attribute its narrow resonant structures to
the superposition of Raman transitions between light-shifted
sublevels of the ground atomic state and to atomic recoil processes.
These two contributions have different dependencies on trap
parameters and we show that the former is inhomogeneously broadened.
The strong dependence of the spectra on the probe-beam polarization
indicates the existence of large optical anisotropy of the cold-atom
sample, which is attributed to the recoil effects. We point out that
the recoil-induced resonances can be isolated from other
contributions, making pump-probe spectroscopy a highly sensitive
diagnostic tool for atoms in a working MOT.
\end{abstract}


\maketitle

\section{\label{sec:intro}Introduction}
The magneto-optical trap (MOT) is now a standard tool for obtaining
cold atomic samples. Due to significantly reduced Doppler
broadening, low collision rate and long interaction time, such
samples offer a unique possibility of ultra-precise spectroscopic
measurements. They are especially useful for experiments which
investigate Raman transitions between nearly degenerate energy
levels. The resonant structures in probe spectra observed in cold
media for pump-probe detuning much smaller than the natural
linewidth are due to numerous effects: stimulated Rayleigh
scattering~\cite{lou92} and Raman transitions between either light
shifted sublevels of a ground atomic state~\cite{gri91}, vibrational
energy levels of atoms localized in an optical lattice~\cite{gry01},
or kinetic momentum states of unbound atoms~\cite{gry94}. However,
the MOT is not an ideal tool for systematic investigation of these
effects. Because of the fixed three-dimensional geometry of the
trapping beams, the presence of inhomogeneous trapping magnetic
field, and limited flexibility in varying the trap parameters, it is
difficult to selectively address the phenomena mentioned above. This
is why the MOT usually serves only as an initial stage of
preparation of a cold sample which, after switching off the MOT's
optical and magnetic fields, is subject to more precisely controlled
experimental conditions, such as dedicated pump beam geometry and
polarization~\cite{lou92},\cite{gry94},\cite{ver92},\cite{chen01},\cite{card00},
phase-stabilized pump beams~\cite{hem93}, etc.

Despite these difficulties, cw pump-probe spectroscopy of trapped
atoms performed while the MOT is working, using the trapping beams
as pump light, has been studied by several authors. Such experiments
were first carried out by Tabosa et al.~\cite{tab91} and Grison et
al.~\cite{gri91}. Also, there exists spectroscopic evidence of
optical lattices in a MOT with a special, phase-shift-insensitive,
trapping beam geometry~\cite{schad91}. These experiments indicated
the potential of probe spectroscopy for cold-atom diagnostics.
However, the complexity of such spectra from a working MOT resulted
in their possible application for MOT diagnostics being neglected.
In this paper we revisit the problem and describe our systematic
studies of such spectra.

The aim of this paper is to examine the Raman processes which occur
in a MOT. Their interplay makes the MOT spectra far more complex than
those obtained under simpler experimental conditions. The systematic
study of the MOT spectra is necessary for their possible application
for MOT diagnostics which, as we show below, can be successfully
performed by pump-probe spectroscopy. Working MOT spectroscopy can
constitute a powerful, non-destructive diagnostic tool. It can
provide information on the atomic cloud density~\cite{lez01} and
average Rabi frequency~\cite{zach02}. Below, we concentrate on the
additional possibility of spectroscopic velocimetry of atoms in a
working MOT, based on the recoil-induced resonances (RIR). This
method has been already successfully used to determine the
temperature of the cold
sample~\cite{gry94},\cite{mea94},\cite{dom01},\cite{fisch01} but
under conditions in which RIR constituted the only relevant
contribution to the spectra. In a working MOT, however,
recoil-induced resonances appear accompanied by the other Raman
processes which necessitates clear identification and more thorough
analysis of their individual contributions.

Below, in Sec.~\ref{sec:setup} we briefly characterize our setup
and in Sec.~\ref{sec:results} present in more detail our
experiment and results. We perform spectroscopic measurements of
$^{85}$Rb atoms in a standard MOT~\cite{raab87}. With the MOT's
optical and magnetic fields still turned on, we simultaneously
record two signals: a probe absorption and a four-wave mixing
signal. The latter is generated by atoms as a result of their
nonlinear interaction with the probe and trapping beams and in
such a geometry propagates oppositely to the probe beam
direction~\cite{boyd92}. The spectroscopic signals were acquired
for various trap parameters, e.g., the trapping beams'
intensity and detuning from atomic resonance and the
gradient of the MOT's magnetic field. This allowed us to carry out
a systematic investigation of the phenomena
that affect the shapes of the absorption and
four-wave mixing spectra.

In Sec.~\ref{sec:interpretation}, we show that the theoretical
explanation of the observed spectra has to include two
contributions: the Raman resonances between light-shifted Zeeman
sublevels of the $^{85}$Rb $^2$S$_{1/2}$ ($F=3$) ground state, or
for short, the Raman-Zeeman resonances
(RZR)~\cite{gri91},~\cite{cour93} and the recoil-induced
resonances (RIR)~\cite{gry94},~\cite{ver96},~\cite{guo92}.
Additionally, in order to reach good agreement between the theory
and the experimental data, one has to consider realistic physical
conditions in the MOT, namely, the light intensity and polarization
gradients, the three-dimensional geometry of the trapping beams, and
the inhomogeneity of the quadrupole MOT's magnetic field.
In Sec.~\ref{sec:interpretation} we present interpretation of the
main individual mechanisms that can be recognized in the MOT. We
also note that the difference between the $\sigma^+$ and
$\sigma^-$ absorption spectra is the signature of the optical
anisotropy of the cold atomic sample in a MOT and is due to recoil
resonances induced by the probe and different trapping beams.
Finally, we suggest a method to eliminate the RZR contribution,
thus opening the possibility of real-time,
nondestructive MOT temperature measurement based
on observation of the width of RIR contribution~\cite{gry94} and
conclude our work in Sec.~\ref{sec:conc}.

\section{\label{sec:setup}Experimental Setup}

We use a standard magneto-optical trap~\cite{zach98} in a
stainless-steel vacuum chamber with anti-reflection coated windows.
Our laser setup consists of four home-built diode lasers. One of
them, equipped with an external cavity, is frequency stabilized
using saturated absorption in a rubidium vapor cell, or
alternatively the Doppler-free dichroism method~\cite{was02}. This
laser serves as the master for injection seeding into the trapping
and probe lasers. The 780~nm light from a 70-mW trapping diode laser
is divided into three beams, which have Gaussian radius
$\sigma\approx0.4$ cm and peak intensity
$I_\text{max}=12$~mW/cm$^2$. They are retro-reflected after passing
the trap cell. The frequency shift of the trapping and probe beams
is controlled by several acousto-optic modulators (AOMs). The
typical detuning, $\Delta=\omega-\omega_0$, of the trapping beams of
frequency $\omega$ from the trapping transition resonance frequency
$\omega_0$ is $-3\Gamma$, where $\Gamma=2\pi\times5.98$~MHz is the
transition's natural linewidth. Before injection into the probe
laser diode, the master beam double passes the AOM, which allows us
to tune the frequency $\omega_\text{pr}$ of the probe laser in the
range of $\pm$40 MHz around the master laser frequency. In the
present experiment the probe sweep range was $\pm$3~MHz. Since both
trapping and probe lasers are seeded with the same beam, they are
phase locked. Thus, when these two lasers work in the pump-probe
configuration they provide a spectroscopic resolution which allows
one to distinguish resonant structures with widths of a few kHz,
limited by residual phase fluctuations. The repumping beam is
derived from an independent, free running laser diode, and is
overlapped with one of the trapping beams. The quadrupole magnetic
field is generated by a pair of anti-Helmholtz coils, which produce
an axial gradient of 16 G/cm. Stray dc magnetic fields are zeroed by
three orthogonal pairs of Helmholtz coils. We trap about $10^7$
atoms in the cloud with a Gaussian radius of $\sigma\approx0.9$~mm.
The temperature of our sample, measured by the time-of-flight
method~\cite{brz02}, is about 100~$\mu$K.
\begin{figure}
\includegraphics[scale=0.4]{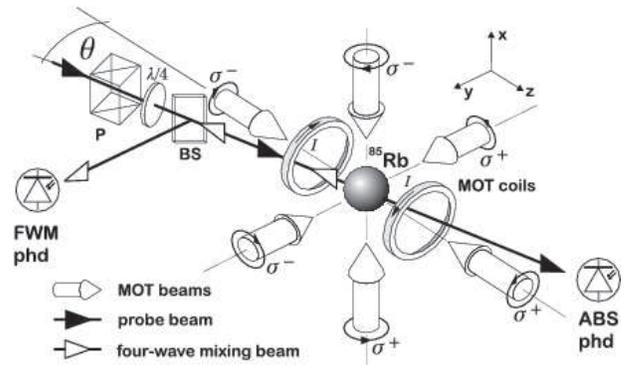}
\caption{\label{fig01}The geometry of the experimental setup.}
\end{figure}
The geometry of our experiment is depicted in Fig.~\ref{fig01}.
The probe beam enters the cloud of cold atoms at
a small angle $\theta=3.5^{\circ}$ with one of the
trapping beams. The polarization of the probe beam is set by the
$\lambda$/4 waveplate placed after a polarizer (P). We define the
polarization of the probe beam with respect to the nearly
co-propagating trapping beam, which is $\sigma^-$-polarized. We
call the probe polarization $\sigma^-$ ($\sigma^+$) when it is the
same as that of the nearly co-propagating (counter-propagating)
trapping beam. After traversing the cloud, the probe beam is
directed onto a photodiode which records the absorption spectrum
(ABS phd). The four-wave mixing beam, generated in the cloud and
propagating oppositely to the probe beam, is reflected by a 50/50
beamsplitter (BS) onto another photodiode (FWM phd). Both signals
are acquired simultaneously and thus can be directly compared. The
probe beam is shaped to have the diameter smaller than the cloud
size in order to avoid an undesired background of
non-absorbed light. Typically, the probe-laser power is about
1~$\mu$W, and its frequency is swept at a rate of 5~MHz/s. The
spectra presented in this paper are recorded as a function of the
pump-probe detuning, $\delta=\omega_\text{pr}-\omega$, and are
averaged over 20 probe sweeps.

\section{\label{sec:results}Measurements and results}

\begin{figure}
\includegraphics[scale=0.4]{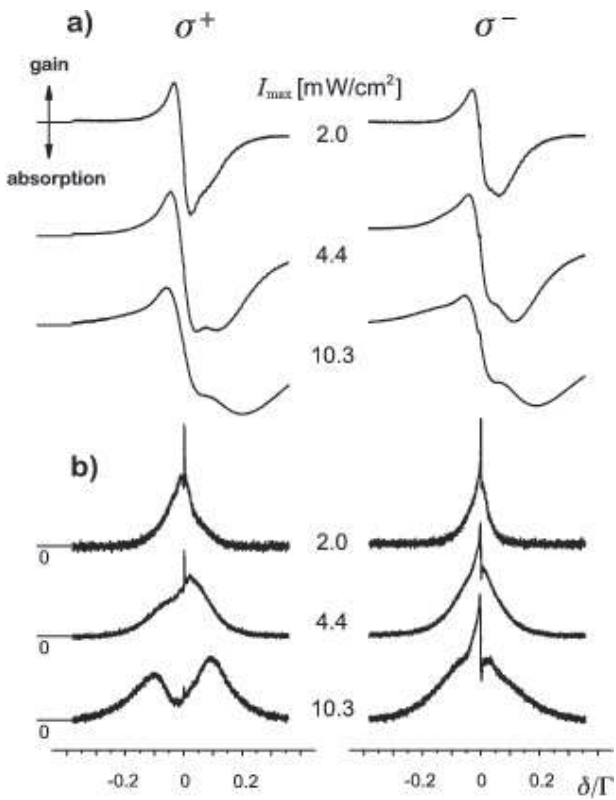}
\caption{\label{fig02}Probe absorption spectra (a) and four-wave
mixing signals (b) for the \sgp{} and \sgm{}-polarized probe beam.
The spectra are recorded as a function of the pump-probe detuning
$\delta$. $I_\text{max}$ stands for the peak intensity of a single
trapping beam. The trapping beam detuning from atomic resonance
is $\Delta=-3\Gamma$, the magnetic field gradient along coil axis
is $\partial B/\partial z=13$~G/cm.}
\end{figure}

We have performed systematic studies of absorption and four-wave
mixing spectra of $^{85}$Rb atoms in a MOT. We have varied the
trapping beams' intensity, their detuning and the
magnetic field gradient. Examples of
absorption and four-wave-mixing spectra simultaneously recorded
for three various trapping beam intensities are depicted in
Fig.~\ref{fig02}. The shapes of the spectra are similar to those
previously observed and assigned to RZR (see, for
example,~\cite{gri91},~\cite{tab91}). However, in contrast to the
pure RZR case, they differ significantly for \sgp{} (left column)
and \sgm{} polarization of the probe beam (right column). In
particular, in absorption, the \sgp{} and \sgm{}  spectra have
different amplitudes and positions of individual resonances. In
four-wave mixing, the spectra consist of two very distinct
contributions, one broad and second ultra-narrow. The broad
contribution to the wave-mixing spectra recorded with the \sgp{} probe
appears broader that that associated with the \sgm{} probe.

From Fig.~\ref{fig02} it is clearly seen that with the increase of
the trapping beam intensity, both the absorption and four-wave-mixing
spectra become wider and their resonant structures better resolved.
This fact can be qualitatively explained by the increase of the
splitting of the ground-state Zeeman sublevels due to the ac Stark
shift (light shift). For $\Omega{}>\Gamma$, this shift is
proportional to the pump beam intensity ($\Omega$ is the Rabi
frequency associated with the pump field). The quantitative
description of the shape of the spectra, however, should incorporate
all possible Raman transitions between the pairs of adjacent Zeeman
sublevels in the $^2$S$_{1/2}$ ($F$=3) ground state of $^{85}$Rb (see
Fig.~\ref{fig04}a below) and also include a possible RIR
contribution. This will be discussed in the next section.

\begin{figure}
\includegraphics[scale=0.4]{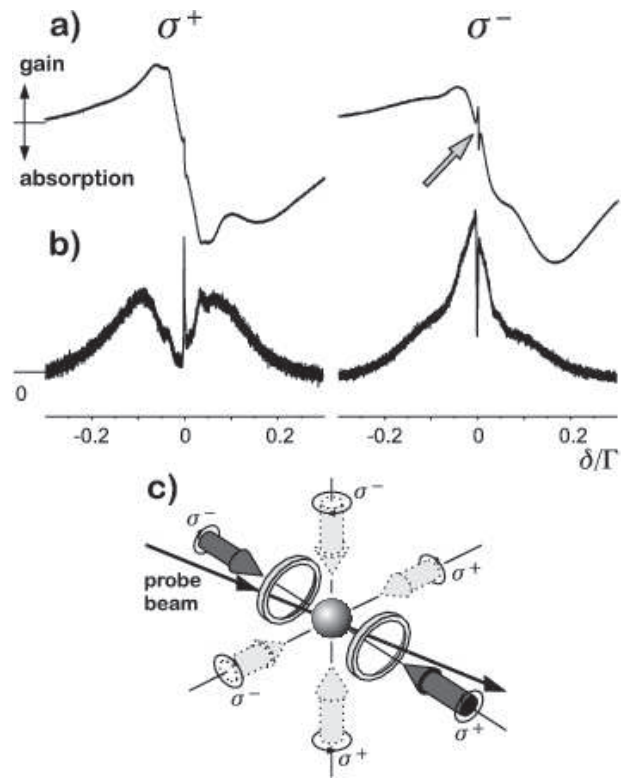}
\caption{\label{fig03}Probe absorption spectra (a) and four-wave
mixing signal (b), recorded when almost all trapping beam power is
sent to the pair of trapping beams nearly collinear with the probe,
as schematically shown in (c). The intensity of these beams is
$I_z=30 \,\,\text{mW/cm}^2$, intensity of beams transversal to the
probe beam is $3\,\,\text{mW/cm}^2$. The detuning of the trapping
beams is $\Delta=-3\Gamma$ and the magnetic field gradient is
$\partial B/\partial z=13$~G/cm. The dispersive shape of the
recoil-induced resonance shows up in the center of the
\sgm{}-polarized probe absorption spectrum (marked by grey arrow).}
\end{figure}

In the case of the absorption spectra for the \sgm{}-polarized
probe, a weak, ultra-narrow structure appears near the pump-probe
detuning $\delta\approx0$. In the four-wave mixing signal this
ultra-narrow resonance is even more pronounced and occurs for both
polarizations of the probe beam. This feature in a MOT spectra has
not yet been thoroughly examined. We attribute it to recoil-induced
resonances~\cite{gry94}. To verify this hypothesis we have performed
an additional measurement. The trapping beam intensities were set in
such a way that almost all trapping laser power was sent to the pair
of trapping beams nearly collinear with the probe. In such a
situation, trapping beams transverse to the probe are much
attenuated and serve only to sustain the cloud stability. Thus, we
approach the one-dimensional pump-probe spectroscopy setup with two
counter-propagating strong pump beams nearly collinear with the
probe~\cite{ver92}, while still having cold atomic cloud in a stable
MOT. Spectra registered under such conditions, presented in
Fig.~\ref{fig03}, reveal resonant structures around $\delta\approx0$
even better resolved than those of Fig.~\ref{fig02}. In particular,
a distinct dispersion-like resonance develops in the center of the
absorption spectrum for the \sgm{}-polarized probe, while for the
\sgp{} probe polarization it is barely visible. The shape of this
resonance agrees well with the theory of recoil-induced
resonances~\cite{gry94} and its width corresponds to the actual
temperature of the atomic sample, which has been measured
independently~\cite{brz02}.

In order to understand how the measured spectra depend on
realistic MOT conditions we have measured the absorption and
four-wave mixing spectra for various magnetic-field gradients.
Since the size of the cloud is proportional to $(\partial
B/\partial z)^{-1/2}$~\cite{raab87}, the value of the magnetic
field in the peripheries of the cloud scales as $(\partial
B/\partial z)^{1/2}$. For a well-aligned MOT with the
value $B=0$ in the trap center, we observed that for smaller
gradients the resonances occur at the same frequencies as for
larger gradients but are better resolved. Thus, due to the finite
size of the atomic cloud, when modelling the experimental curves
one has to take into account the inhomogeneous broadening due to
the MOT quadrupole magnetic field.

\section{\label{sec:interpretation}Interpretation}

The shape of the absorption and four-wave mixing spectra that we
measure in our experiment can be explained on the basis of two
processes: Raman transitions between light-shifted Zeeman sublevels
(RZR) of the ground state~\cite{gri91},~\cite{cour92} and
recoil-induced resonances (RIR)~\cite{gry94}. The theoretical
outline of both processes is presented below. However, in order to
perform the complete modelling of the spectra, we extended the
present theories by inclusion of two types of inhomogeneous
broadening: one due to modulation of the net \textbf{E}-field in the
trap and the other connected with the presence of the quadrupole
trapping magnetic field \textbf{B}. Moreover, the standard
three-dimensional geometry of the MOT beams has to be taken into
account when considering recoil processes in a working trap.

\subsection{\label{subsec:RZR}Raman transitions between light-shifted Zeeman sublevels (RZR)}

\begin{figure}
\includegraphics[scale=0.3]{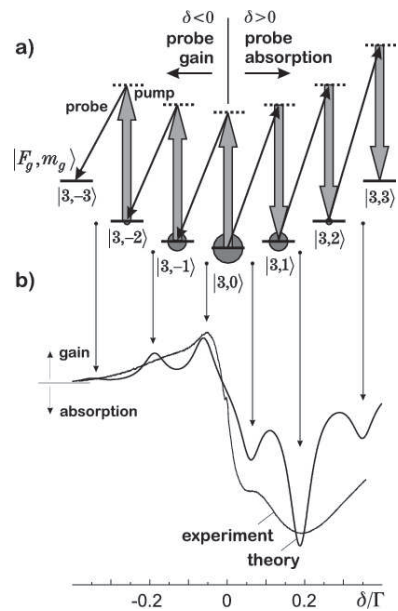}
\caption{\label{fig04}(a) Multilevel structure of the 5
$^2$S$_{1/2}$ ($F=3$) $^{85}$Rb ground state. Zeeman sublevels are
perturbed with a $\pi$-polarized pump (thick grey arrows), which
results in their energy shift and symmetric population distribution
(alignment), marked by grey circles. The structure is then probed by
a \sgp{}-polarized beam. (b) Theoretical absorption signal compared
with the experimental spectrum. The theoretical curve was generated
for the Rabi frequency $\Omega =6\Gamma$, detuning $\Delta
=-3\Gamma$ and $\gamma_{i,k}=0.13\Gamma$ for all $i$, $k$.
Experimental conditions: $I_\text{max}=10.2 \,\,\text{mW/cm}^2$,
$\Delta =-3\Gamma$, $\partial B/\partial z=13$~G/cm.}
\end{figure}

\begin{figure}
\includegraphics[scale=0.25]{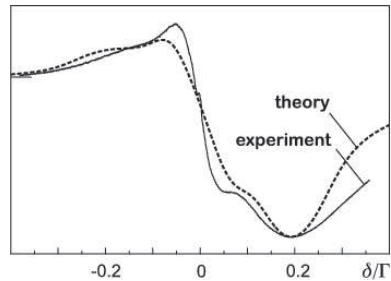}
\caption{\label{fig05}The comparison of experimental absorption
spectrum (acquired under the same experimental conditions as in
Fig.~\ref{fig04}b) with the theoretical curve generated according to
(\ref{eq:SRZR}) and averaged over the Rabi frequency range
$\Omega\in[5\Gamma;\,7.5\Gamma]$.}
\end{figure}

Let us consider the cycling 5 $^2$S$_{1/2}$ ($F_g = 3$) $-$ 5
$^2$P$_{3/2}$ ($F_e = 4$) transition in $^{85}$Rb atom in a MOT. The
atom is subject to three pairs of counter-propagating laser beams of
frequency $\omega{}$ with pairwise orthogonal circular
polarizations. Their interference in the intersection region results
in a complex spatial modulation of light intensity and
polarization~\cite{hop97}. For the sake of simplicity, we consider
here only the case where the net light field is linearly polarized
and choose quantization axis parallel to the local direction of the
net field \textbf{E}. In this reference frame, the resulting pump
light is $\pi$-polarized and shifts the Zeeman sublevels of both the
ground and the excited atomic state. The 5 $^2$S$_{1/2}$ ($F_g=3$)
ground state multilevel structure is presented in Fig.~\ref{fig04}a.
The optical pumping in such a scheme leads to alignment, i.e. a
symmetric distribution of populations with respect to the $m_g=0$
sublevel of the ground state, with this sublevel being mostly
populated. The atoms interacting with the pump light are probed by
circularly polarized weak probe laser~\cite{comm01}. The Raman
processes involving a $\pi$-polarized pump and a
$\sigma^\pm$-polarized probe photon lead to transitions with $\Delta
m_g=\pm1$. For a given population distribution among the $m_g$
sublevels, two directions of such processes are possible, depending
on the sign of the probe-pump detuning $\delta$. In the considered
case of dominant population in the $m_g=0$ sublevel, the Raman
transitions with $\Delta m=-1$ take place for $\delta < 0$ and lead
to gain of the probe; those with $\Delta m=+1$ take place for
$\delta>0$ and result in its attenuation. These processes are
resonant whenever $|\delta|$ coincides with the energy separation of
the adjacent sublevels. The amplitude of the corresponding resonance
is proportional to the population difference of the sublevels
involved in the transition and depends on the Clebsch-Gordan
coefficients associated with the specific transition path. The
simplest model of the probe absorption spectrum is obtained by
summing the Lorentzian profiles centered at the appropriate
resonance frequencies and weighted by products of the relevant
population differences and squares of the Clebsch-Gordan
coefficients
\begin{eqnarray}
\nonumber
s_\text{RZR}(\delta)=\sum_{i=1}^{3}w_{i-1,i}\Delta\Pi_{i-1,i}L\left(\delta,\delta_{i-1,i},\gamma_{i-1,i}\right)\\
\label{eq:SRZR}
-\sum_{i=-1}^{-3}w_{i+1,i}\Delta\Pi_{i+1,i}L\left(\delta,\delta_{i+1,i},\gamma_{i+1,i}\right).
\end{eqnarray}
In the above equation, $w_{k,j}$ is the weight associated with the
Clebsch-Gordan coefficient along the $k=m_g \leftrightarrow j=m_g$
Raman transition path, $\Delta\Pi_{k,j}$ is the population
difference of the $k=m_g$ and $j=m_g$ sublevels, and the
Lorentzian profile is given by
\begin{equation}
\label{eq:L}
L\left(\delta,\delta_{k,j},\gamma_{k,j}\right)=\frac{\gamma_{k,j}}{\gamma_{k,j}^2+(\delta-\delta_{k,j})^2},
\end{equation}
where $\delta_{k,j}$ and $\gamma_{k,j}$ are, respectively, the
resonance frequencies and widths of the $k=m_g \leftrightarrow
j=m_g$ transitions. These widths are due to the optical pumping, the
finite interaction time and atomic collisions in a trap. The
spectrum generated using formula (\ref{eq:SRZR}) with assumption of
full optical pumping in the closed system shown in Fig.~\ref{fig04}a
is presented in Fig.~\ref{fig04}b compared to the relevant
experimental data. The comparison shows that the latter constitutes
an envelope for the theoretical curve under which individual
resonances are unrealistically well resolved. This suggests a
broadening mechanism, which is indeed provided by the spatial
inhomogeneity of the light intensity in the trap. To account for
this effect, we averaged the calculated spectra over Rabi frequency
$\Omega$. This can be done as described in the Ref.~\cite{marq96},
but in our case it was sufficient to sum the absorption spectra
generated according to Eq.~\ref{eq:SRZR} for the appropriate range
of Rabi frequencies. By numerical simulations of various field
contributions, similar to that of Ref.~\cite{hop97}, we have found
that the relevant range that produces the best agreement between the
theory and the experiment is $5\Gamma$ to $7.5\Gamma$. This
corresponds to the average Rabi frequency of $6.25\Gamma$ determined
as in Ref.~\cite{zach02}. The inclusion of the \textbf{E}-field
inhomogeneity yields a better agreement of theoretical results with
the experimental data, as presented in Fig.~\ref{fig05}. However,
there is still a discrepancy between the theoretical prediction and
the experimental signal near $\delta\approx0$. Moreover, the model
discussed above does not explain the profound difference between the
spectra registered for \sgp{} and \sgm- polarized probe. The
explanation of this effect is presented in the next subsection.

\subsection{\label{subsec:RIR}Recoil-induced resonances (RIR)}

The momentum exchange between non-localized atoms and the laser
field is associated with the phenomenon of the so-called
recoil-induced resonances (RIR). They were predicted by Guo et
al.~\cite{guo92} and observed by Grynberg et
al.~\cite{gry94},~\cite{ver92}. A simple momentum-space analysis
of RIR can be found in
Refs.~\cite{gry94},~\cite{fisch01},~\cite{ver96}. Here we briefly
recall its results.

\begin{figure}
\includegraphics[scale=0.3]{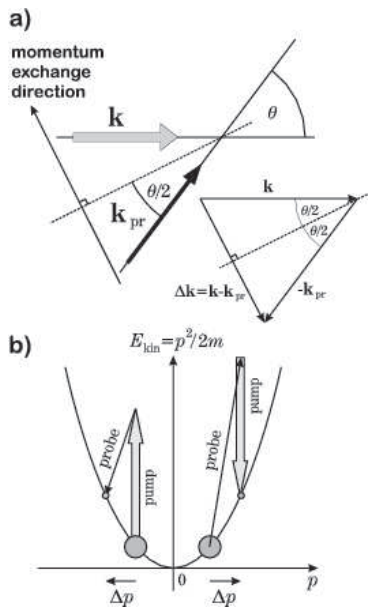}
\caption{\label{fig06}(a) Basic geometry for the recoil induced
resonance, associated with atomic interaction with two beams: the
pump (thick grey arrow) and probe (thin arrow) beams intersecting in
an atomic sample at a small angle $\theta$. (b) Raman transitions
between the kinetic momentum states for an atom in a state with
specific $m_g$.}
\end{figure}

Let us consider atoms of mass $m$ interacting with a weak probe
beam (with wave vector $\mathbf{k}_\text{pr}$ and frequency
$\omega+\delta$), which makes an angle $\theta$ with the direction
of propagation of a strong pump beam (wave vector $\mathbf{k}$,
frequency $\omega$), as depicted in Fig.~\ref{fig06}a. The process
of pump photon absorption followed by the probe photon emission
results in the momentum change $\Delta p$ of an atom, where
\begin{equation}
\label{eq:Dp} \Delta p=-2\hbar k \sin\frac{\theta}{2}.
\end{equation}
The preceding equation is derived assuming small probe-pump detuning
$|\mathbf{k}|\approx|\mathbf{k}_\text{pr}|=k$. The resonance
occurs whenever the probe-pump detuning
coincides with the kinetic energy difference, namely at detuning
\begin{equation}
\label{eq:dres} \delta_\text{res}=-\frac{2k}{m}\left(\hbar k
\sin{\frac{\theta}{2}}-p\right)\sin{\frac{\theta}{2}}.
\end{equation}
An analogous consideration performed for the
case of probe-photon absorption followed by emission of
a photon into the pump beam leads to the nearly
identical formula for $\delta_\text{res}$ as (\ref{eq:dres}), but
with the opposite sign. Similarly to the RZR case, the resonance
amplitude is proportional to the population difference between the
relevant atomic states, but now we consider the
continuum, kinetic momentum states rather than discrete, magnetic
sublevels (Fig.~\ref{fig06}b),
\begin{equation}
\label{eq:DeltaP}
\Delta\Pi\left(p_\text{final},p_\text{initial}\right)=\Pi\left(p+\Delta
p\right)-\Pi\left(p\right).
\end{equation}
Integration over all possible momentum values leads to the formula
for the RIR signal given as
\begin{equation}
\label{eq:sRIR}
s_\text{RIR}(\delta,\theta)=-\int_{-\infty}^{\infty}\mathrm{d}p\Delta\Pi\left(p_\text{final},p_\text{initial}\right)L(\delta).
\end{equation}
Assuming that $\Pi(p)$ is a Gaussian distribution, $\Delta p \ll
p_T=\sqrt{mk_BT}$ and $\gamma\ll kp_T/m$~\cite{gry94} (all these
conditions are fulfilled in our case), one arrives at an analytic
formula for the absorption spectrum signal
\begin{eqnarray}
\nonumber
s_\text{RIR}(\delta,\theta)&=&-\sqrt{\frac{m}{2\pi}}\frac{\hbar
\delta}{2u^{3/2}_Tk\sin{(\theta/2})}\\ \label{eq:sRIRa} & & \times
\exp\left[-\frac{\delta ^2}{2u^2_T(2k\sin(\theta/2))^2}\right],
\end{eqnarray}
where $u_T = p_T /m$ is the most
probable atomic speed. The signal (\ref{eq:sRIRa}) is the derivative
of a Gaussian function. It has its
minimum/maximum for $\delta=\pm 2u_T k\sin(\theta /2)$ and the
width of the spectrum $\Delta_{RIR}$, defined as the distance
between minimum and maximum, is proportional to $\sqrt{T}$. The
RIR spectrum can thus serve as a spectroscopic tool for the
temperature measurement of a cold atomic
sample~\cite{gry94},\cite{mea94},\cite{dom01}, according to the
formula
\begin{equation}
\label{eq:temp} T=\frac{m}{16k_Bk^2\sin(\theta/2)}\Delta_{RIR}^2.
\end{equation}
The results presented above refer to a two-level atom and a simple
pump-probe beam configuration. Nevertheless, this simple approach
can be also successfully used in our case in which multilevel
atoms are subject to a three dimensional set of trapping
beams. Since a distinct RIR contribution appears only
near $\delta\approx0$ (see Figs.~\ref{fig02},~\ref{fig03}), it
must result from Raman
transitions between the momentum states of the same magnetic
sublevel, $\Delta m_g=0$. In principle, RIRs can also occur
between momentum states associated with various Zeeman sublevels,
i.e. with $\Delta m_g\neq 0$, in which case some \sgp/\sgm{}
asymmetry might be visible, but this contribution to the overall
signal is negligible.
First, the amplitudes of such contributions have been shown to be
about order of magnitude weaker than of the $\Delta m_g=0$
principal ones~\cite{guo93}. Second, they would result only in a
small frequency shift of the broader RZR contribution. Third, as
shown below in~\ref{subsec:Bfield}, the Raman transitions with
$\Delta m_g\neq 0$ in a working MOT are subject to inhomogeneous
broadening which further reduces their importance for RIR. Thus,
the recoil contribution is due to the probe and pump beams of the
same polarization. This implies that our atoms can be effectively
treated as two level systems~\cite{lez01},~\cite{lip00} and that
the polarization of the probe selects the pump of the appropriate
polarization among all six available beams.

For example, consider the case in which the
probe beam is \sgm{}-polarized and the quantization axis is
parallel to its direction of propagation. Obviously, one
contribution to the recoil process is due to the probe beam
combined with the nearly co-propagating trapping beam.
Analogously, the \sgp{}-polarized probe interacts with the nearly
counter-propagating trapping beam. The second contribution is due
to the trapping beams transverse to
the probe. In the chosen reference frame, these beams appear as
linearly $\sigma$ and $\pi$-polarized. Depending on their relative
phases, they represent \sgp{} or \sgm{}-polarized pump photons
with the same probability, so they equally contribute to the
recoil processes both for the \sgp{} and \sgm{}-polarized probe.
This consideration yields the total RIR signal in the probe beam
absorption in the MOT with all six trap beams as equal to
\begin{equation}
\label{eq:sRIRMOT} s_\text{RIR,\,MOT}(\delta)=\left\{
\begin{aligned}s_\text{RIR}&(\delta,\theta)+2s_\text{RIR}(\delta,90^{\circ})\\& \text{for \sgm{}-polarized probe;}\\
s_\text{RIR}&(\delta,180^{\circ}-\theta)+2s_\text{RIR}(\delta,90^{\circ})\\&
\text{for \sgp{}-polarized probe.}
\end{aligned}\right.
\end{equation}
The above discussion allows one to associate the observed
distinct difference between the MOT spectra taken for \sgm{} and
\sgp{}-polarized probe exclusively with the recoil processes,
since, as it was pointed out in Sec.~\ref{subsec:RZR}, the RZR
contributions are insensitive to the probe beam polarization.

\subsection{\label{subsec:Bfield}Influence of the MOT magnetic field}

\begin{figure}
\includegraphics[scale=0.4]{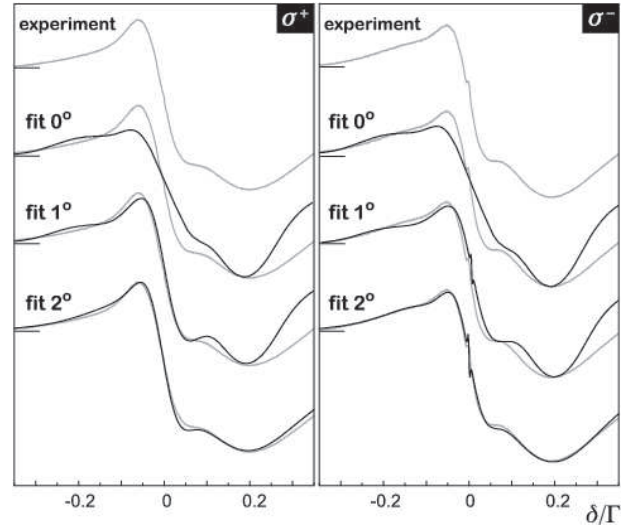}
\caption{\label{fig07}Experimental and theoretical absorption spectra
for \sgp{} and \sgm{}-polarized probe beam. The experimental
conditions: $I_\text{max}=10.2 \,\,\text{mW/cm}^2$, $\Delta
=-3\Gamma$, $\partial B/\partial z=13$~G/cm. Fit $0^\circ$ includes
only inhomogeneity of the $\mathbf{E}$-field, fit $1^\circ$ - as fit
$0^\circ$ + RIR, fit $2^\circ$ - as fit $1^\circ$ + inhomogeneity of
the $\mathbf{B}$-field. Parameters of theoretical simulations: Rabi
frequency averaging range $\Omega\in[5\Gamma;\,7.5\Gamma]$; Gaussian
width of a single, inhomogeneously broadened resonance
$\sigma_B=0.088\Gamma$, which corresponds to the actual trap size
$\sigma_z =0.9$ mm; the temperature of atomic cloud $T=155\,\mu$K was
fitted to achieve the best agreement of the simulation with
experimental data and is consistent with the results of
Ref.~\cite{brz02}. For better comparison, the experimental spectra
are added in grey to each fit.}
\end{figure}

The considerations presented above lead to the conclusion that
both RZR and RIR contribute to the observed spectra, so the
complete formula for the absorption signal becomes
\begin{equation}
\label{eq:scomplete}
s(\delta)=\bar{s}_\text{RZR}(\delta)+s_\text{RIR,MOT}(\delta),
\end{equation}
where $\bar{s}_\text{RZR}(\delta)$ is the RZR contribution
averaged over Rabi frequencies available in a MOT, as discussed in
Sec.~\ref{subsec:RZR}. The results of theoretical simulations
according to formula (\ref{eq:scomplete}) are depicted in
Fig.~\ref{fig07} (fit $1^\circ$) and compared with experimental
data. While the agreement of the modelled and experimental spectra
improved considerably after considering both contributions to
$s(\delta)$, the theoretical spectrum still exhibits resonances
that are too narrow. Simplistic attempts to reduce resolution of the
theoretical spectra by extending the $\Omega$ inhomogeneity range
in the $\bar{s}_\text{RZR}(\delta)$ contribution does not improve
the fit. Whereas the net spectrum $s(\delta)$ becomes broader, its
individual peaks and dips are shifted out of coincidence with the
experimental ones. This indicates that some other kind of
inhomogeneous broadening mechanism must affect the spectra. This
mechanism is due to the quadrupole MOT magnetic field $\mathbf{B}$
and finite size of the atomic cloud. As the spatial distribution
of atomic density in the trap is Gaussian, modelling of the
individual RZR resonances, such as in Eq.~(\ref{eq:SRZR}),
should be performed with the Gaussian
profiles $G(\delta,\delta_{k,j},B)$, rather than with the Lorentz
functions (\ref{eq:L}). We thus take
\begin{equation}
\label{eq:G}
G(\delta,\delta_{k,j},B)=C\exp\left[-\frac{\left(\delta-\delta_{k,j}\right)^2}{2\sigma_B^2}\right],
\end{equation}
with
\begin{equation}
\label{eq:sigmaB} \sigma_B=\frac{g_F\mu_B}{\hbar}\frac{\partial
B}{\partial z}\sigma_z\gg\gamma_{k,j}.
\end{equation}
In the above formulas, $C$ is a constant, $g_F$ is the Lande
factor of the $^2$S$_{1/2}$($F=3$) hyperfine state, $\mu_B$ is the
Bohr magneton and $\sigma_z$ is the Gaussian radius of the atomic
cloud along $z$ axis.
The results of the theoretical simulation of the absorption spectra
with inclusion of the MOT magnetic field are presented in
Fig.~\ref{fig07} as fit $2^\circ$ and are compared with the
experimental data. In contrast to averaging exclusively over
$|\mathbf{E}|$ (fits $0^\circ$ and $1^\circ$), the agreement of the
theoretical and experimental absorption signals is now very good.
This indicates the importance of the inhomogeneous broadening of the
MOT spectra. So far, the RZRs in a MOT have been considered to be
free of any inhomogeneous broadening. Our experiment shows that this
is not the case. While Doppler broadening is negligible, the
magnetic field inhomogeneities cause substantial broadening of the
pump-probe Raman spectra and corresponding change of their
lineshapes from Lorentzian to Gaussian.

\subsection{\label{subsec:FWM}Four-wave mixing signals}

In our experiment, the four-wave mixing spectra are recorded
simultaneously with the probe beam absorption, which assures that
they are acquired in exactly the same experimental conditions. The
four-wave mixing signals recorded in our experiment
(Figs.~\ref{fig02}b,~\ref{fig03}b) clearly exhibit two
contributions: the broad pedestal and pronounced, ultra-narrow
features for $\delta\approx 0$. The former can be attributed to the
superposition of the RZRs and RIRs and the latter is due to the
transitions between kinetic momentum states, RIR and possibly to the
Rayleigh scattering~\cite{lou92}. Ultra-narrow resonances are here
far better resolved than in the case of corresponding absorption
signals. Hence, the four-wave mixing spectroscopy appears to be more
sensitive to the recoil effects. However, precise verification of
this conjecture is hampered by the level of complexity of the theory
of four-wave mixing for recoil induced resonances (see,
e.g.~\cite{guo92}). Also, the nature of the four-wave mixing process
differs substantially from the probe absorption: the four wave
mixing signal is, in general, calculated as $|\rho_{eg}(p,p\,')|^2$,
where $\rho_{eg}(p,p\,')$ is an off-diagonal element of a
momentum-dependent density matrix. Since
$\rho_{eg}(p,p\,')=\rho_{eg}(p,p\,')_\text{RZR}+\rho_{eg}(p,p\,')_\text{RIR}$,
the interference terms
$\rho_{eg}(p,p\,')_\text{RZR}\times\rho_{ge}(p\,',p)_\text{RIR}$
play an important role in the four-wave mixing signal. These terms
are likely to be responsible for significant differences between
four-wave mixing signals for the two circular probe polarizations.
The theoretical work by Guo et al.~\cite{guo92} presents a
derivation of the four-wave mixing resonances due to atomic recoil
in a two-level system. However, for this theory to be applicable to
our case, it should be complemented with the RZR contribution for
systems with nonzero angular momenta.

\subsection{\label{subsec:optan}Optical anisotropy of cold atoms}

\begin{figure}
\includegraphics[scale=0.3]{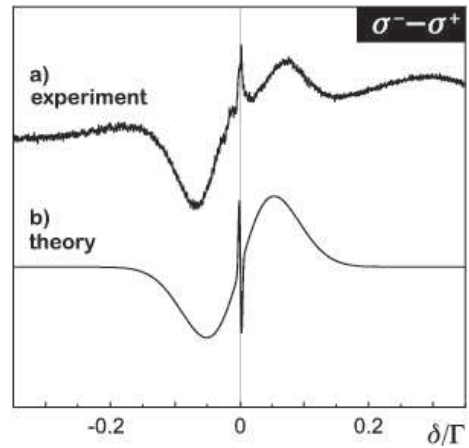}
\caption{\label{fig08}The difference \sgm$-$\sgp{} for the
experimental absorption spectra presented in Fig.~\ref{fig07} (a)
compared with theoretical prediction (b).}
\end{figure}

In Sec.~\ref{subsec:RZR} we discussed the RZR contribution to the
MOT absorption spectra. It does not depend on the probe beam
polarization, due to the fact that in the considered case of
$\pi$-polarized pump field, neither populations of individual Zeeman
sublevels, nor the Raman-resonance frequencies, depend on the sign
of $m_g$. On the other hand, in the case of RIR, the polarization of
the probe beam plays a crucial role. As was mentioned in the
conclusions of Sec.~\ref{subsec:RIR}, recoil processes for two
opposite circular polarizations of the probe result in dramatically
different momentum and energy transfers. The momentum transfer from
trapping beams perpendicular to the probe is the same for both probe
beam polarizations and, according to Eq.~(\ref{eq:Dp}), equals
$\Delta p=\pm\hbar k \sqrt{2}/2$. However, the probe beam
polarization selects which of the two pump beams, the co-propagating
or the counter-propagating with the probe, is involved in the RIR.
The quantity of the momentum transfer is $\Delta p\approx \pm\hbar k
\theta$ for the \sgm{}- and $\Delta p\approx \pm2\hbar k$ for the
\sgp{}-polarized probe. Hence, the RIR contribution is the cause of
the obvious difference between the absorption spectra for the two
probe beam polarizations around $\delta\approx 0$. Fig.~\ref{fig08}a
exemplifies such a difference, obtained by subtraction of two
independently measured absorption signals, compared with the
calculated one (Fig.~\ref{fig08}b). While wide structures of the two
curves (between -0.1 $\Gamma$ and +0.1 $\Gamma$) are well
correlated, the agreement of their central parts is less
satisfactory. The exact cause of this discrepancy is not yet well
understood. There are several possible effects that can influence
that challenging recording of kHz-wide spectral features. On one
hand the accuracy with which we determine the zero of the pump-probe
detuning is limited, the relative phase of the two laser beams
undergoes some residual fluctuations and the MOT parameters may vary
between successive acquisitions of the $\sigma^{-}$ and $\sigma^{+}$
spectra. On the other hand, additionally to the RZR and RIR
contributions that we consider, the central structure might be
systematically affected by additional mechanism such as, e.g. the
Rayleigh scattering~\cite{lou92} or Faraday rotation due to
imperfect compensation of a magnetic field and/or imperfect
balancing of the MOT beams.

The asymmetry between absorption spectra, i.e. between optical
properties of the sample as seen by probe beams of different
polarizations, is known as \textit{dichroism}, or generally as
\textit{optical anisotropy}. In our case, the atomic sample is
placed in a center of a MOT, where $B=0$, and the  optical pumping
by pump beams creates atomic alignment, rather than polarization.
Thus the optical anisotropy of our cold sample can be associated
exclusively with the recoil processes. Consequently, when two
absorption spectra for the \sgm{}  and \sgp{}  probe beams are
subtracted, the RZR contribution is completely eliminated and one
is left only with the contribution from RIRs for the pump beams
co- and counter-propagating with the probe.

The optical anisotropy associated with RIR and shown above
(Fig.~\ref{fig08}) constitutes experimental evidence of the effect
recently predicted by Dubetsky and Berman~\cite{dub95}. The
authors of this paper concentrated on the
somewhat different experimental situation in
which the optical anisotropy is detected via rotation of
the polarization plane of a weak, linearly polarized
probe beam propagating perpendicularly to the pump
beam~\cite{comm02}. Still, the anisotropy described here has the
same physical origin, the RIR.

\section{\label{sec:conc}Conclusions}

We have successfully explained absorption spectra recorded in a
working MOT. We have shown that full agreement between the theory
and experimental data is attainable only when realistic MOT
conditions, the light modulation in a trap and the quadrupole MOT
magnetic field are taken into account. Our theoretical analysis
emphasizes the importance of the atomic recoil processes in a MOT.
Recoil-induced resonances have already been used in atomic
velocimetry~\cite{mea94},\cite{dom01}. In order to apply this method
to atoms in a working MOT, one has to eliminate other processes
influencing the spectra, namely the Raman transitions between
light-shifted Zeeman sublevels of the ground atomic state, RZR. For
the MOT velocimetry based on RIRs, the RZR contribution appears as
an undesired background which hinders precise determination of the
RIR width. Unlike the RIR contribution, that due to the RZR is
subject to significant inhomogeneous broadening by trap magnetic and
light fields, so it is worthwhile to eliminate this background, e.g.
such as suggested above by recording the optical anisotropy or the
differential \sgp{}/\sgm{} absorption.

Let us note that in contrast to the experiment by Schadwinkel et
al.~\cite{schad91} with carefully phase-stabilized trap beams, we
see no contribution from the Raman transitions between vibrational
levels associated with a possible optical lattice, whose
frequencies could be in the similar range of the order of 100~kHz.
We have found that under regular conditions in a standard MOT with
nonstabilized trap beams, the short-term, random
fluctuations of the trapping beam phases caused by mechanical
instabilities of the setup, wash out the interference pattern,
thus preventing stationary modulation of optical potential and
atomic localization.

Another aspect of our results is the four-wave mixing
spectroscopy. Since the four-wave mixing signals are generated in
a more complex process, especially when the multilevel structure
and recoil effects have to be taken into account, their
theoretical analysis is complicated and requires further
investigation. At the same time they appear to be far more
sensitive to recoil processes than absorption, as the ultra-narrow
structures are here highly pronounced. This interesting feature
motivates one to work out the method of MOT diagnostics based on
four-wave mixing.

\subsection*{\label{sec:ackn}Acknowledgments}

This work was supported by the KBN (grant no.~2P03B~088~26) and is
part of a general program on cold-atom physics of the National
Laboratory of AMO Physics in Torun, Poland. We are grateful to
T.~Pałasz for his contribution to the early stage of the experiment,
A.~Gabris for his help in numerical calculations, M.~Trippenbach and
D.~Budker for stimulating discussions and to D.~Kimball, Sz.~Pustelny,
and S.~Rochester for their comments on the manuscript.


\begin{thebibliography}{99}
\bibitem{comm01}In our model, we consider the characteristic $\pi-\sigma^\pm$ pump-probe polarization configuration, which is the simplest one that
can reproduce the shape of the observed spectra.
\bibitem{comm02}In Ref.~\cite{dub95}, the term Faraday rotation is used for the predicted rotation associated with RIR, though it is not due to any
magnetic field. Some authors distinguish the magnetic-field from otherwise induced rotations by using the notions of "paramagnetic" and "diamagnetic"
Faraday effects~\cite{coh67}. Similarly as in the recent review~\cite{bud02}, we reserve the notion of the Faraday effect for a specific narrow class
of effects exclusively related to the presence of the magnetic field and prefer to use the term optical anisotropy for the broad class of cases where
the rotation is not related to the magnetic field, as in the case discussed in this paper.
\end{thebibliography}
\end{document}